\documentclass[conference]{IEEEtran}
\usepackage{cite}
\usepackage{booktabs}
\usepackage{array}
\usepackage{siunitx}
\usepackage{amsmath,amssymb,amsfonts}
\usepackage{algorithmic}
\usepackage{enumitem} 
\usepackage{multirow}
\usepackage{graphicx}
\usepackage{textcomp}
\usepackage{xcolor}
\def\BibTeX{{\rm B\kern-.05em{\sc i\kern-.025em b}\kern-.08em
    T\kern-.1667em\lower.7ex\hbox{E}\kern-.125emX}}
\begin{document}

\title{LiDAR-Aided Millimeter-Wave Range Extension using a Passive Mirror Reflector}


\makeatletter
\newcommand{\linebreakand}{%
  \end{@IEEEauthorhalign}
  \hfill\mbox{}\par
  \mbox{}\hfill\begin{@IEEEauthorhalign}
}

\author{\IEEEauthorblockN{Omar Ibrahim,  Raj Sai Sohel Bandari,  and Mohammed E.  Eltayeb}
\IEEEauthorblockA{Department of Electrical and Electronic Engineering, California State University,  Sacramento,  Sacramento,  USA\\
Email: \{omaribrahim2, rbandari, mohammed.eltayeb\}@csus.edu}
}

\maketitle

\begin{abstract}
Passive reflectors mitigate millimeter-wave (mm-wave) link blockages by extending coverage to non-line-of-sight (NLoS) regions.  However, their deployment often leads to irregular reflected beam patterns and coverage gaps.  This results in rapid channel fluctuations and potential outages.  In this paper,  we propose two LiDAR-aided link enhancement techniques to address these challenges.  Leveraging user position information,  we introduce a location-dependent link control strategy and a user selection technique to improve NLoS link reliability and coverage.  Experimental results validate the efficacy of the proposed techniques in reducing outages and enhancing NLoS signal strength. 
\end{abstract}


\section{Introduction}

Millimeter-wave (mm-wave) communication has emerged as a promising technology to address the rising demand for low latency and high-speed  wireless networks \cite{p2, p3}.   Nonetheless,  the  dependence on line-of-sight (LoS) links and sensitivity to blockages limits the effective range and reliability of mm-wave systems \cite{choles,PR2}.  This problem is exacerbated in dynamic and dense indoor environments where intermittent blockages of LoS links are imminent.   Addressing mm-wave link blockages and extending coverage to beyond-LoS regions are therefore critical prerequisites for the realization of high-speed reliable mm-wave networks.

Various solutions in the literature explored the use of passive metallic and non-metallic reflectors to extend mm-wave coverage to non-line-of-sight (NLoS) regions and mitigate the blockage problem \cite{PR2 , 2,  PR3, PR4, PR5}.  These solutions exploit the specular and diffuse signal scattering properties of reflecting surfaces to redirect mm-wave signals around obstacles without amplifying or modifying them. This presents a cost-effective alternative to emerging reconfigurable intelligent surfaces that typically require a power source and additional channel resources for reflection beam training  \cite{LiS1, LiS2}.   Despite their effectiveness,  passive reflectors create non-uniform reflection patterns and spotty coverage  leading to inefficiencies and  potential communication outages  \cite{PR2}.  Many factors influence the reflection pattern including the propagation environment,  size of the reflector,  distance from the reflector,  and the beam pattern of the transmit and receive antennas.  Therefore,  NLoS channel state and user location information are necessary for effectively adapting the mm-wave communication link and mitigating potential outages.

In this paper, we investigate the potential of combining Light Detection and Ranging (LiDAR) with a flat mirror reflector to enhance mm-wave communication coverage within an L-shaped corridor as illustrated in Fig. \ref{fig:IR}. Utilizing LiDAR-derived user position data,  we introduce two passive reflector communication enhancement techniques.  The first technique employs a location-dependent received signal strength (RSS) back-off approach to reduces outages in the NLoS area.  The second technique opportunistically allocates the mm-wave channel to user(s) in favorable coverage areas.  We show that user location information is instrumental in enhancing reflector coverage,  optimizing resource utilization,  and ensuring reliable beyond-LoS mm-wave connectivity.

 \section{Measurement Set-Up and Scenario}

\subsection{Measurement Environment}
MM-wave measurements were conducted in an L-shaped corridor as illustrated in Fig. \ref{fig:IR} and Fig. \ref{fig:lidar_mmwave}.  The receiver was strategically positioned at different locations within the (x, y) plane of the corridor to establish a receiver grid. These measurements were performed under standard non-line-of-sight conditions, with silver, copper, and mirror reflectors placed at the corridor's periphery. The transmitter height was 1.5 meters and 3.8 meters away from the reflector within the 2.5-meter width corridor. The receiver height was 1.5 meters,  and the RSS was measured over 102 grid points  where each gird dimension was $0.3 \times 0.3$ m$^2$ as shown in  Fig. \ref{fig:lidar_mmwave}.  A foam board at angle $45^o$ was set up at the corner to support the reflecting surfaces placed at 1.35 m above the ground.

\subsection{Measurement Hardware}
The measurement hardware consisted of two 60 GHz mm-wave phased antenna array transceiver kits (Sivers EVK02001),  a passive reflector,  a universal software radio peripheral  (USRP),  laptops,  and a  Quanergy  M-8 LiDAR sensor \cite{LiDAR}.  The mm-wave phased antenna arrays  were employed for  transmission and reception.  The Quanergy  M-8 LiDAR was used to obtain user distance measurements and tracking for situational awareness.  The transmit beam was steered towards the reflector at an angle of departure of $-5^o$ to maximize coverage,  while the receive beam was fixed at angle of arrival of $0^o$, i.e.  towards the reflector. The beam pattern of the employed antenna arrays can be found in \cite{Deepsense}.

\begin{figure}
\center
    \includegraphics[width=0.8\linewidth]{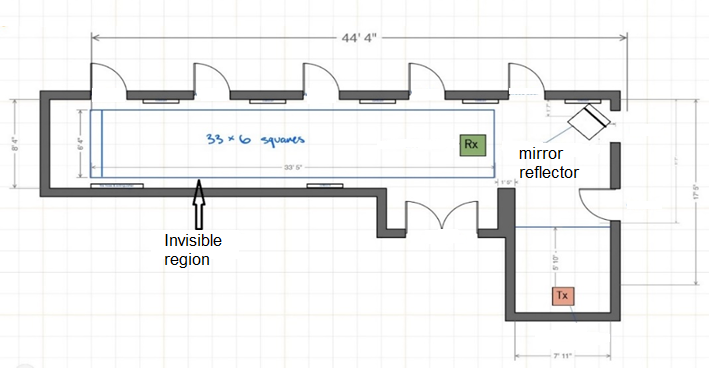}
    \caption{Illustration of the indoor measurement environment with a reflecting surface deployed at the corner of the corridor.}
    \label{fig:IR}
\end{figure}

 \begin{figure}
 \center
    \includegraphics[width=0.65\linewidth]{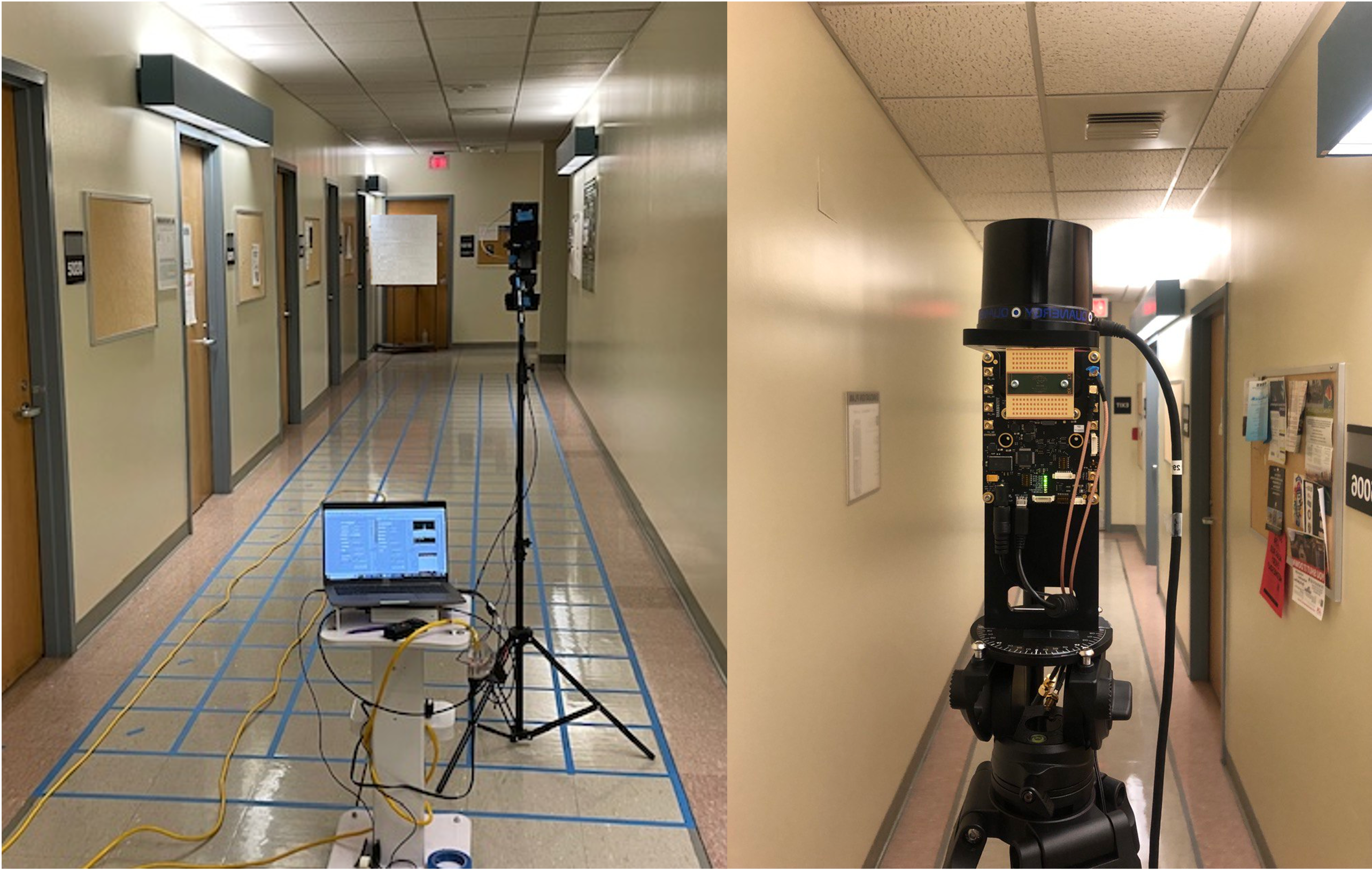}
    \caption{Left: View of the L-shaped corridor featuring 102 grid points for RSS measurements,  the receiver,  and the flat foam board.  Right: image of the collocated mm-wave transmitter and LiDAR  at the end of the corridor.} 
    \label{fig:lidar_mmwave}
\end{figure}

\section{LiDAR-Assisted mm-wave Range Extension using Passive Mirror Reflector}
\subsection{LiDAR Sensing in NLoS Environments}
LiDAR has become an indispensable technology in various fields  for capturing environmental information and improving situational awareness.   However, its field of view (FoV) is often limited by physical environmental obstructions that can occlude the LiDAR's view.  To overcome this limitation,  this research explores the use of mirrors to redirect the LiDAR's laser beam  to NLoS areas and enable beyond-LoS sensing.   We demonstrate this in Fig.  \ref{fig:fig2} where a mirror is used to reflect LiDAR signals around an L-shaped corridor.  Fig.  \ref{fig:fig2} shows successful tracking of a user walking towards the L-shaped corner occlusion confirming the mirror's ability to  expand the LiDAR's field of view.  By integrating a LiDAR sensor and a millimeter-wave transmitter on the same platform and employing mirror reflectors, we can create a shared field of view. This allows for simultaneous LiDAR sensing and millimeter-wave communication within an expanded NLoS area.

\subsection{Mirror Reflector for  MM-Wave Range Extension}
Mirrors have long been employed in interior design to mitigate blind spots and optimize visibility by redirecting light around obstacles. This principle can be extended to mm-wave signals. By strategically placing mirrors in the environment, it's feasible to expand mm-wave  coverage and overcome LoS limitations.  Typically composed of glass and a metallic coating, mirrors effectively reflect mm-wave signals due to the reflective properties of both materials. Nonetheless, the choice of mirror material significantly influences mm-wave reflection characteristics. Metals like copper, aluminum, and gold excel in mm-wave conductivity, making them suitable reflectors.  Conversely, glass introduces attenuation due to absorption and penetration losses, which results in lower signal strength in NLoS areas when compared to  metal-only reflectors. Building on these insights, this paper explores the potential of silver-coated mirrors, combining silver and glass, for reflecting both mm-wave and LiDAR signals for joint mm-wave communication and LiDAR sensing range extension.

\begin{figure}
 \center
    \includegraphics[width=0.65\linewidth]{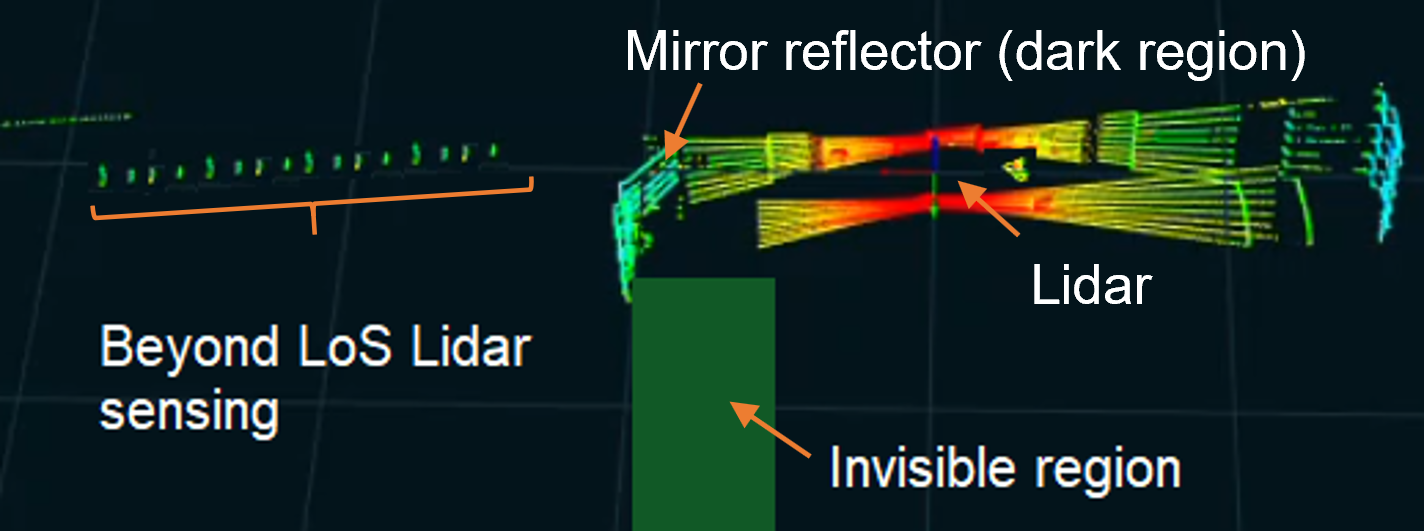}
\caption{ LiDAR tracking of a user (1.8 m tall) in an L-shaped corridor using a mirror reflector.  Mirror assisted LiDAR enables NLoS localization with centimeter level accuracy. }
    \label{fig:fig2}
\end{figure}

\subsection{Proposed Coverage Enhancement Techniques}
Passive reflectors effectively extend coverage into targeted NLoS regions.  Nonetheless, these reflectors induce inconsistent power levels within the NLoS coverage, resulting in communication interruptions as users change position.  In the following,  we utilize LiDAR-derived user location data and prior RSS coverage measurements to optimize NLoS coverage.
\subsubsection{Enhanced NLoS Coverage Through Location-Based RSS Control}
To enhance coverage and reduce communication disruptions, we propose a location-aware RSS back-off approach that conservatively underestimates the RSS at the receiver.  Specifically,  let $\gamma_d$ denote the actual RSS at position $d$,  and $\Delta_d$ represent a location dependent back-off constant.  Rather than using $\gamma_d$ as the RSS,  the transmitter assumes a reduced RSS of $\gamma_d-\Delta_d$.  In this paper, we let $\Delta_d = \frac{\kappa}{\bar{\gamma}_\text{nn}}$, where $\kappa$ is a constant and $\bar{\gamma}_\text{nn}$ is the nearest neaighbor average RSS.  This strategy prioritizes back-off in weak coverage areas,  effectively minimizing outages while preserving data rates.  

\subsubsection{RSS Maximization Through User Selection}
Non-uniform reflection patterns create spatially varying RSS levels.  To optimize mm-wave transmission,  the NLoS area is divided into an $N$-point grid where RSS measurements are mapped to LiDAR-derived locations and mm-wave links are only scheduled to users within high-RSS regions.  If a single user is detected in a strong coverage region,  the user is exclusively assigned the mm-wave link.  For multiple users in a high-RSS region,  a random user is selected.  By focusing on optimal coverage regions,  we enhance NLoS link performance and maximize the transmitter's data rate.

\begin{figure}
\centering
    \includegraphics[width=0.55\linewidth]{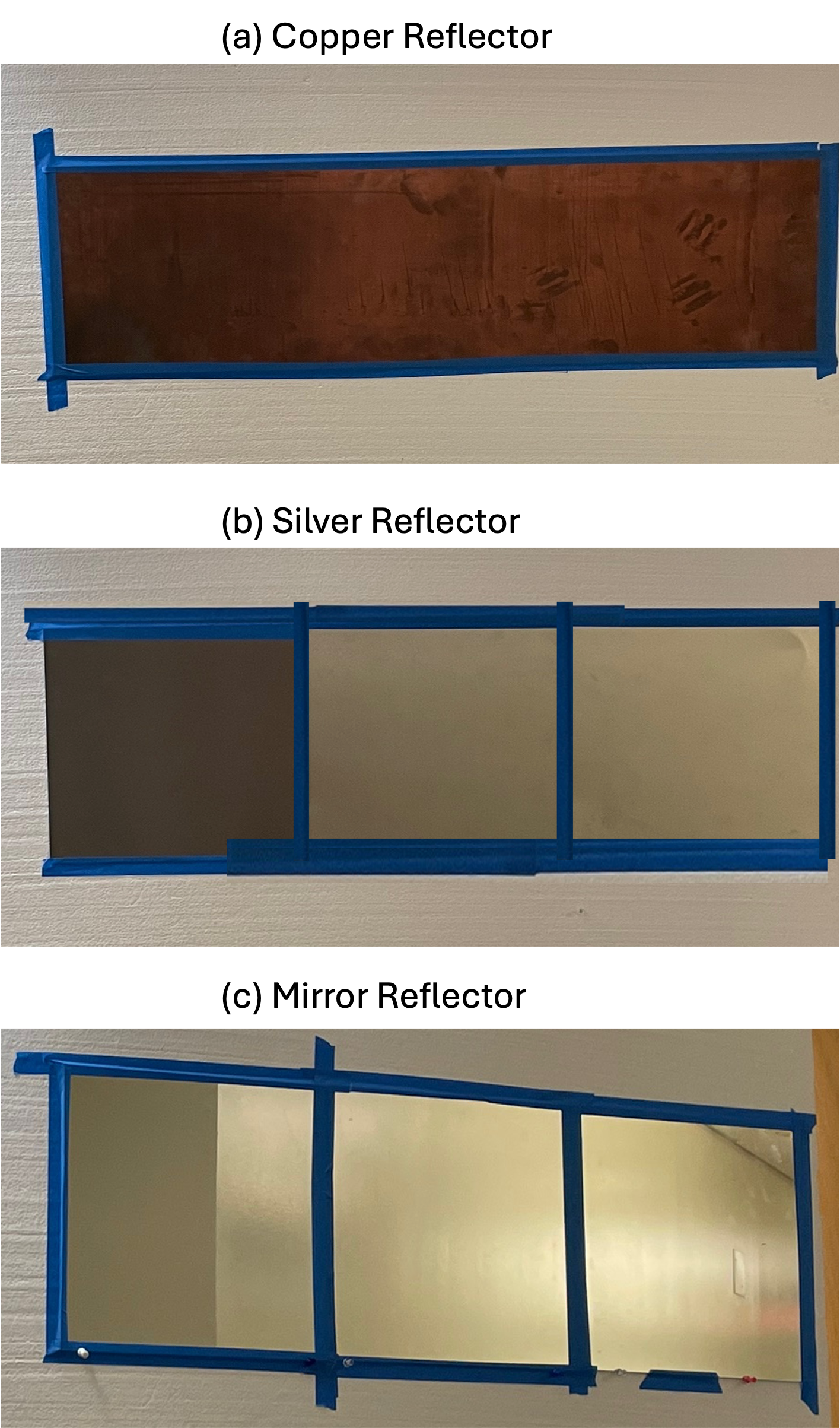}
\caption{ Flat passive reflectors of size $0.3 \times 0.9$ m$^2$ mounted on a foam board at an azimuth angle of $45^o$ with respect to the transmitter.}
    \label{fig:figR}
\end{figure}

\section{Experimental Results and Discussions} 
This section presents a proof-of-concept demonstration of the proposed NLoS link enhancement techniques. To evaluate their performance,  RSS and LiDAR measurements were conducted at 102 grid points within the obstructed area using a silver-coated mirror reflector, in addition to silver and copper reflectors for comparison purposes.  All reflectors were attached to a $1.2 \times 1.2$ meter foam board angled at $45^o$ as shown in Fig. \ref{fig:figR}.  

Table \ref{Table1} summarizes LiDAR detection performance for different mirror sizes.  LiDAR coverage is defined as the proportion of grid points with detected users.  The results  highlight the impact of the mirror size on expanding the LiDAR FoV.  In Fig. \ref{fig:ref}, we present the RSS values measured at each grid point when using different reflector types. The figure shows that all reflectors produced strong overall RSS in the NLoS region but exhibited uneven coverage.  These results are consistent with the RSS measurements in \cite{2}. The foam board, serving as a baseline without a reflector yielded the lowest RSS emphasizing the need of reflectors for NLoS signal improvement.  The corresponding quantitative results are shown in Fig. \ref{fig:ccdf} where we  plot the complementary cumulative distribution function (CCDF) of the total RSS  across the grid for all reflectors types. The CCDF is defined as the probability that the  RSS exceeds an RSS threshold $\gamma_{\text{th}}$, i.e. $\text{Pr} (\text{RSS} > \gamma_{\text{th}})$.  Results indicate that silver reflectors provide the highest likelihood of strong RSS,  equally followed by copper and mirror, with foam exhibiting the lowest performance.  

In Fig. \ref{fig:pout},  we illustrate the relationship between outage probability and user location displacement.  The outage probability is defined as $\text{P}_\text{out} = P(\gamma_{d_0} < \min(\gamma_{d_1}, \gamma_{d_2}, ...,\gamma_{d_i}))$,  where $\gamma_{d_0}$ is the RSS at distance $d_{0}$ from the reflector and $\gamma_{d_i}$ is the RSS at distance $d_i$,  with $|d_0-d_i| >0$.  A MATLAB simulation of 1000 random user locations within the grid was conducted to generate the plot,  assuming user movement towards the reflector. The results indicate outage probability increases with user displacement for all reflector types, with silver exhibiting the highest outage.  The proposed back-off strategy is shown to effectively reduce outages, with larger back-off values correlating to lower outage probabilities.  The corresponding back-off values are visualized in Fig. \ref{fig:bo}, demonstrating higher back-off in low-RSS regions and lower back-offs in high-RSS regions (see Fig.  \ref{fig:ref}).  This strategy preserves RSS levels in high coverage regions.

\begin{figure}
 \center
\hspace{-10mm}
 \includegraphics[width=1.1\linewidth]{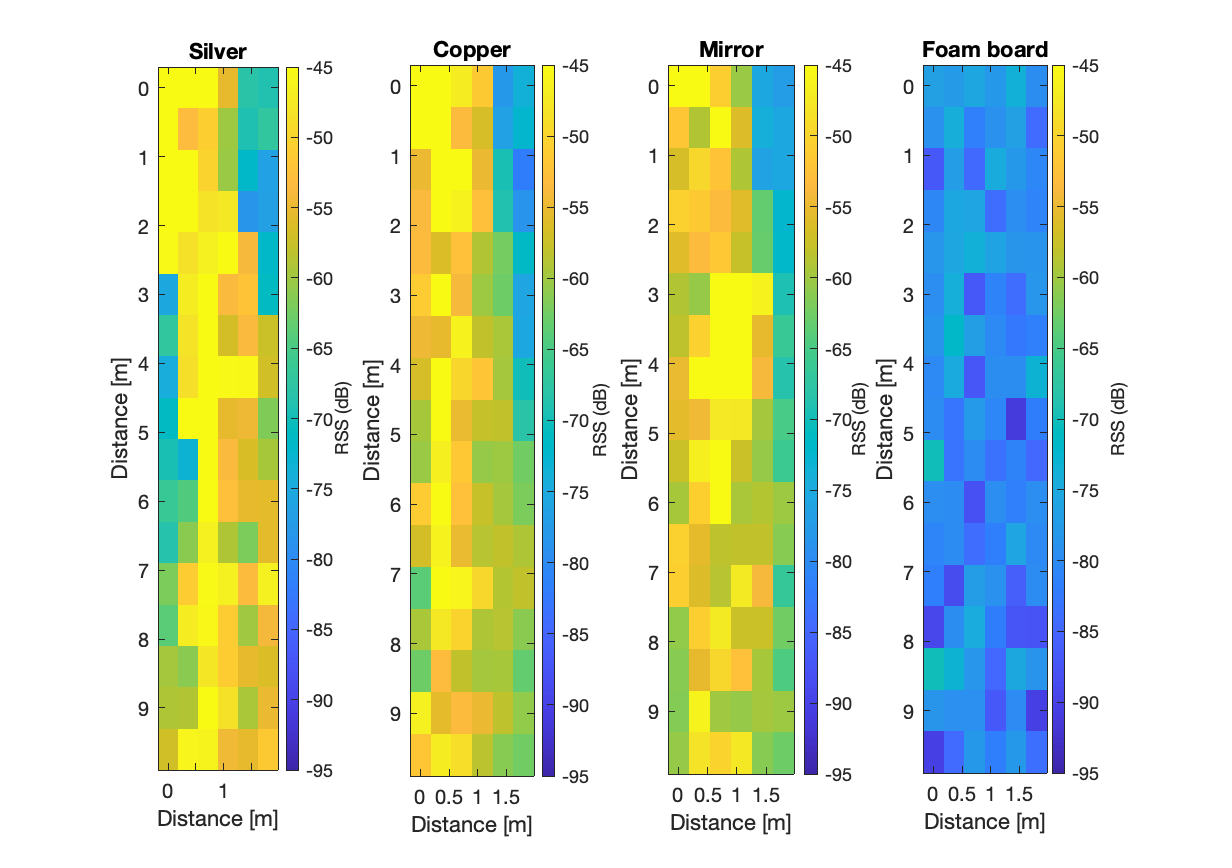}
\caption{Total RSS values on 102 grid points obtained when using a silver reflector,  a copper reflector,  a silver-coated mirror, and a foam board (no reflector attached).  All reflectors are flat panels of size $0.3 \times 0.9$ m$^2$.  }
\label{fig:ref}
\end{figure}

\begin {table}[t] 
\centering \small
    \caption{  LiDAR detection range using varying flat reflector sizes.} 
\begin{tabular}{|c|c|c|}
 \hline
Reflector Size &LiDAR Detection\\
 \hline
Mirror: $0.3  \times  0.9$ $m^2$   & 95\%\\
Mirror: $0.3  \times  0.6$  $m^2$ & 89\%   \\
Mirror: $0.3  \times  0.3$ $m^2$  & 77\%   \\
 \hline
\end{tabular} \label{Table1}
\end{table}

To evaluate the impact of LiDAR-aided user selection on NLoS RSS maximization,  we generated 1000 random instances of $k$ user locations across the grid using MATLAB.   All simulated user locations have direct LoS  to the reflector.  At each instance (channel use),  the RSS of the user in the strongest coverage area was recorded.  The CCDF plots of the RSS values corresponding to different number of users are shown in Fig.  \ref{fig:us}.   Results indicate that leveraging multi-user diversity through user selection enhances NLoS link performance by increasing minimum RSS.

\section{Conclusions and Future Work}
This paper explored the use of mirror reflector for extending both 60 GHz mm-wave coverage and LiDAR FoV in an indoor environment.  Two link enhancements techniques leveraging user location information were proposed.   Results demonstrated the effectiveness of user location information in reducing communication outages and improving the schedued NLoS RSS.  Future work will investigate advanced beamforming techniques that leverage multiple reflectors for multi-user scheduling and interference management,  as well as  coverage optimization through actuator control and LiDAR-based NLoS blockage prediction algorithms.

\begin{figure}
\center
 \includegraphics[width=7.5cm]{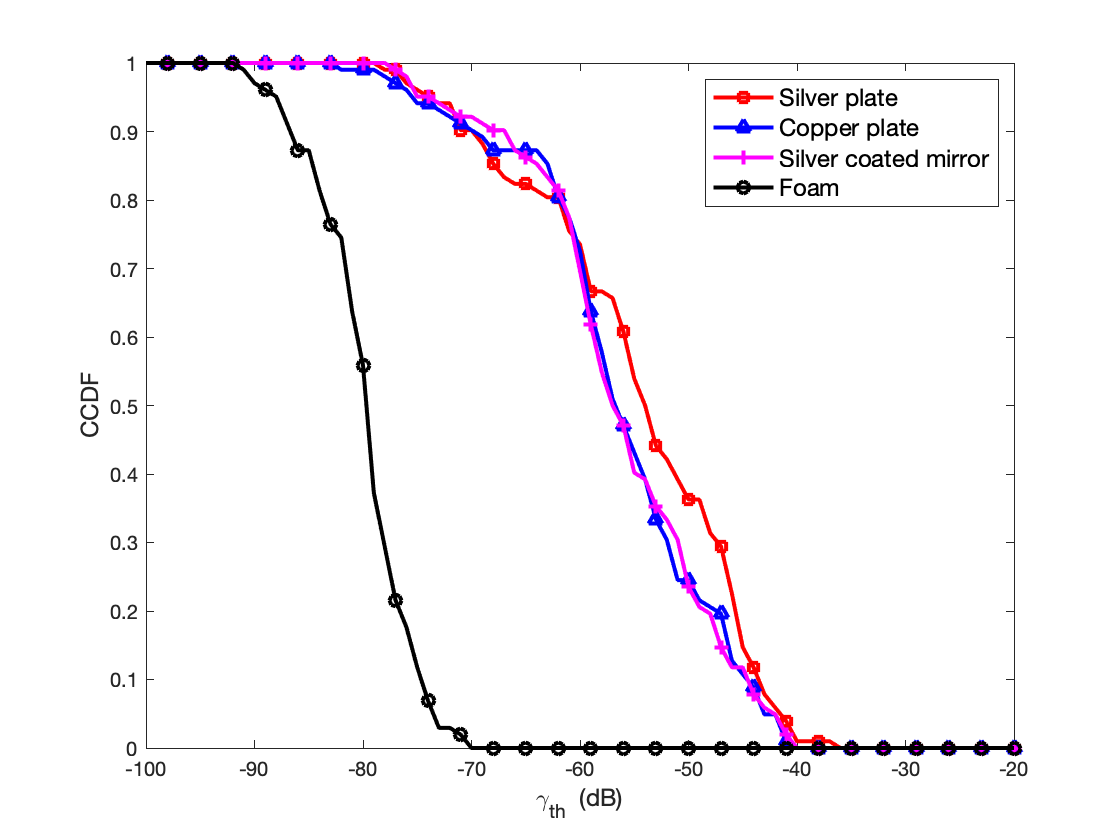}
\caption{Complementary cumulative distribution function of the total RSS values for multiple reflectors types. }
\label{fig:ccdf}
\end{figure}

\begin{figure}
\center
 \includegraphics[width=7.5cm]{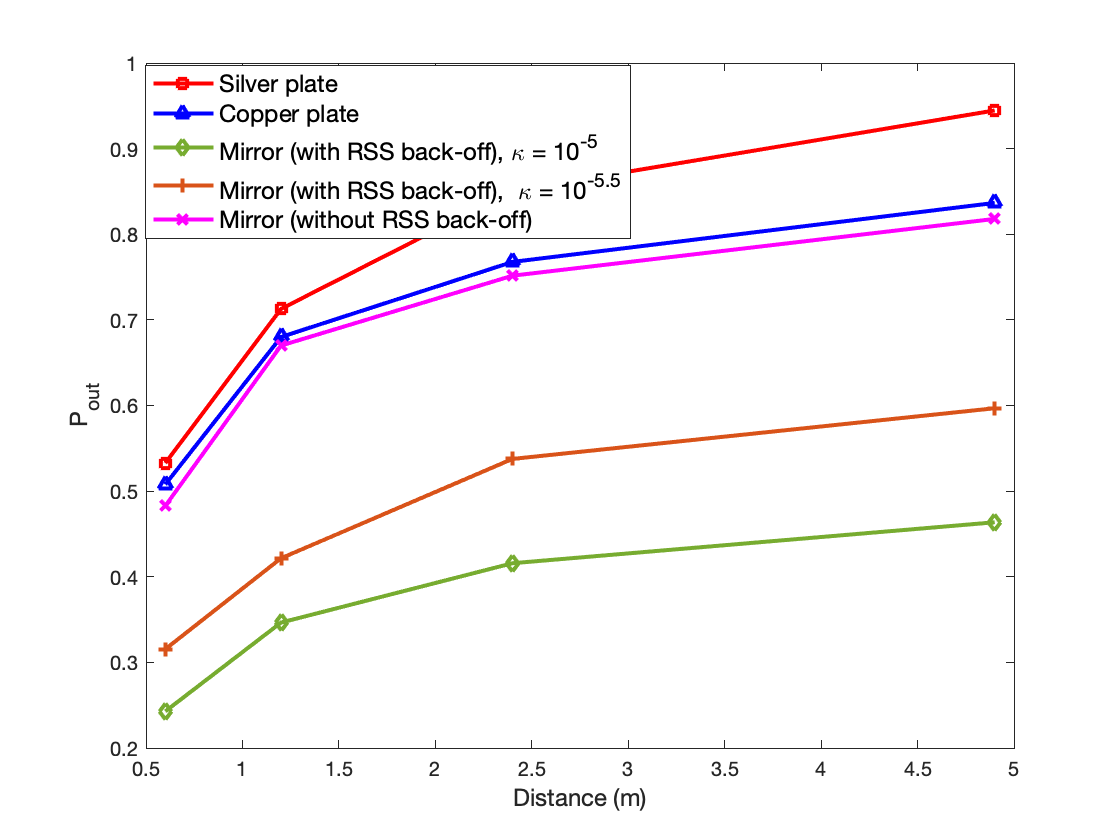}
\caption{Outage probability versus the user displacement (distance).  }
\label{fig:pout}
\end{figure}

\begin{figure}[t]
\center
 \includegraphics[width=7cm]{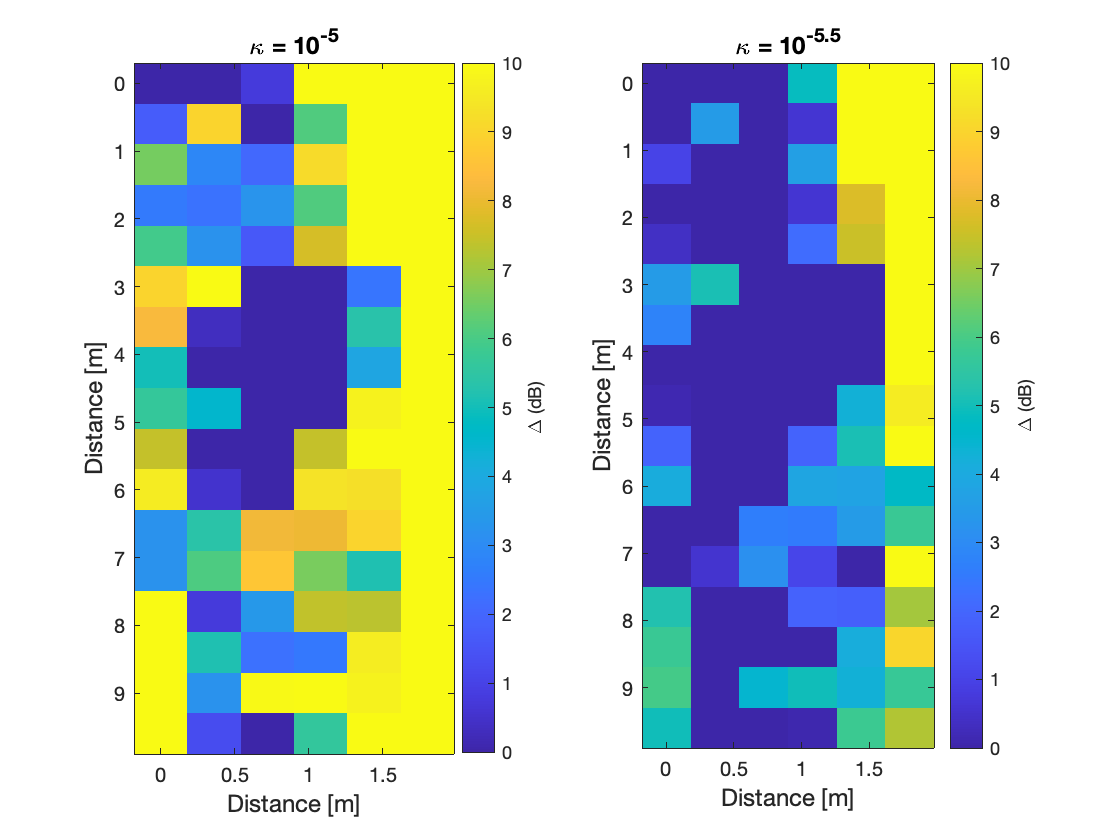}
\caption{Spatial distribution of the RSS back-off values for mirror reflector.}
\label{fig:bo}
\end{figure}

\begin{figure}
\center
 \includegraphics[width=7.8cm]{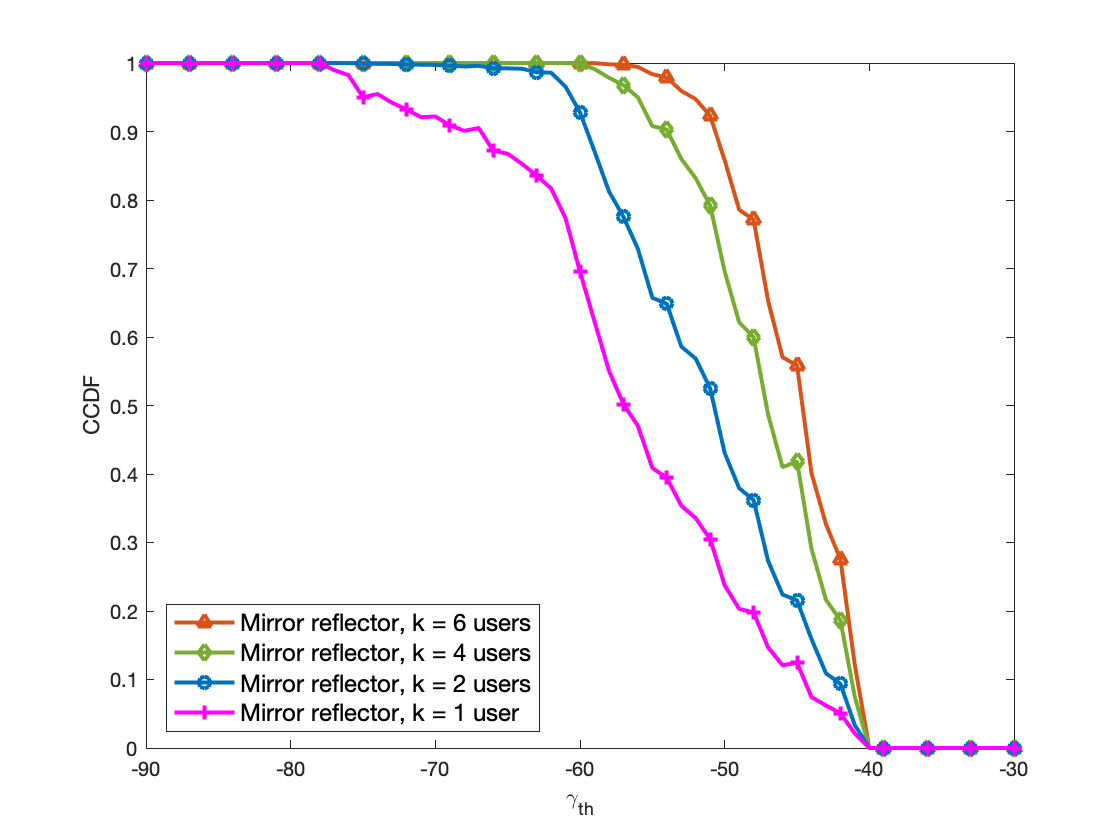}
\caption{Complementary cumulative distribution function of the total RSS values for multiple number of users. }
\label{fig:us}
\end{figure}

\section*{Acknowledgment}
This material is based on work supported by the National Science Foundation under grant No. NSF-2243089.


\begin{thebibliography}{1}



\bibitem{p2} 
J. Zhang et al.,  ``A Survey of mmWave-Based Human Sensing: Technology, Platforms and Applications,''  in \emph{IEEE Communications Surveys \& Tutorials, } vol. 25, no. 4, pp. 2052-2087, Fourthquarter 2023.


\bibitem{p3} 
T. S. Rappaport, Y. Xing, O. Kanhere, S. Ju, A. Madanayake, S. Mandal, A. Alkhateeb, and G. C. Trichopoulos,  ``Wireless communications and applications above 100 GHz: Opportunities and challenges for 6G and beyond,'' IEEE Access, vol. 7, pp. 78 729–78 757, 2019.

\bibitem{choles} 
C. K. Anjinappa, F. Erden and I. Güven,  ``Base Station and Passive Reflectors Placement for Urban mmWave Networks,'' in \emph{IEEE Transactions on Vehicular Technology,} vol. 70, no. 4, pp. 3525-3539, April 2021. 


\bibitem{PR2}
K. Qian, Y. Lulu, Z.  Xinyu, and T. Ng,  ``MilliMirror: 3D printed reflecting surface for millimeter-wave coverage expansion,'' in \emph{ Proceedings of the 28th Annual International Conference on Mobile Computing And Networking,} pp. 15-28, 2022.


\bibitem{2}
      W. Khawaja, O. Ozdemir, Y. Yapici, F. Erden and I. Guvenc,  ``Coverage Enhancement for 
      NLOS mm-wave Links Using Passive Reflectors,''  in \emph{IEEE Open J.  of the  Communications Society,}  vol. 1, pp. 263-281, 2020. 


\bibitem{PR3}
Z. Yu, C. Feng, Y. Zeng, T. Li and S. Jin, ``Wireless Communication Using Metal Reflectors: Reflection Modelling and Experimental Verification,'' \emph{IEEE International Conference on Communications,} Rome, Italy, 2023, pp. 4701-4706.


\bibitem{PR4}
C. K. Anjinappa et al.,``Indoor Propagation Measurements with Transparent Reflectors at 28/39/120/144 GHz,''  \emph{2022 IEEE International Conference on Communications Workshops (ICC Workshops),} Seoul, Korea, Republic of, 2022, pp. 1118-1123. 


\bibitem{PR5}
M. El Hajj, M. Dieng, G. Zaharia and G. El Zein, ``Enhancement Indoor mmWave Coverage Using Passive Reflector for NLOS Scenario,'' in \emph{2022 16th European Conference on Antennas and Propagation (EuCAP),} Madrid, Spain, 2022, pp. 1-5. 

\bibitem{LiS1}
A. Taha, M. Alrabeiah and A. Alkhateeb,  ``Enabling Large Intelligent Surfaces With Compressive Sensing and Deep Learning,'' in \emph{IEEE Access,}  vol. 9, pp. 44304-44321, 2021.


\bibitem{LiS2}
M. Di Renzo et al., ``Smart Radio Environments Empowered by Reconfigurable Intelligent Surfaces: How It Works, State of Research, and The Road Ahead,''  in \emph{IEEE Journal on Selected Areas in Communications, } vol. 38, no. 11, pp. 2450-2525, Nov. 2020.

%





\bibitem{Deepsense} 
A. Alkhateeb,  G. Charan, T. Osman, A. Hredzak,  N. Srinivas,  U. Demirhan,  and J. Morais,  ``DeepSense 6G: A Large-Scale Real-World Multi-Modal Sensing and Communication Dataset,'' \emph{ IEEE Dataport},   October 31, 2022.

\bibitem{LiDAR} 
M.-A. Mittet, H. Nouira, X. Roynard, F. Goulette, and J.-E.  Deschaud,  ``Experimental assessment of the Quanergy M8 lidar sensor,''  in \emph{ ISPRS 2016 congress,} Prague, Czech Republic, July 2016.






                

    
%


   
    
\end{thebibliography}
\end{document}